\documentclass[]{spie}  

\usepackage{amsmath,amsfonts,amssymb}
\usepackage{graphicx}
\usepackage{multirow}
\usepackage[colorlinks=true, allcolors=blue]{hyperref}

\title{The Infrared Imaging Spectrograph (IRIS) for TMT: Data Reduction
System}

\author[a,b]{Gregory Walth}
\author[a,b]{Shelley A. Wright}
\author[c]{Jason Weiss}
\author[c]{James E. Larkin}
\author[d]{Anna M. Moore}
\author[e]{Edward L. Chapin}
\author[c]{Tuan Do}
\author[e]{Jennifer Dunn}
\author[f]{Brent Ellerbroek}
\author[f]{Kim Gillies}
\author[g]{Yutaka Hayano}
\author[c]{Chris Johnson}
\author[a,b]{Daniel Marshall}
\author[d]{Reed L. Riddle}
\author[e]{Luc Simard}
\author[c]{Ji Man Sohn}
\author[g]{Ryuji Suzuki}
\author[d]{James Wincensten}

\affil[a]{Department of Physics, University of California San Diego, CA,
92039, USA;}
\affil[b]{Center for Astrophysics \& Space Sciences, University of
California San Diego, CA, 92039, USA;}
\affil[c]{Physics \& Astronomy Department, University of California Los
Angeles, CA 90095 USA;}
\affil[d]{Caltech Optical Observatories, 1200 E California Blvd., Pasadena,
CA 91125 USA;}
\affil[e]{National Research Council of Canada - Herzberg, Victoria, BC, V9E
2E7 Canada;}
\affil[f]{Thirty Meter Telescope Observatory Corporation, Pasadena, CA
91105 USA;}
\affil[g]{National Astronomical Observatory of Japan, Osawa, Mitaka, Tokyo,
181-8588 Japan}

\authorinfo{Further author information: (Send correspondence to G.L.W.)\\ G.L.W.: E-mail: gwalth@ucsd.edu, Telephone: 1 858 822 3435}

\begin{document}
\maketitle

\begin{abstract}
IRIS (InfraRed Imaging Spectrograph) is the diffraction-limited first light
instrument for the Thirty Meter Telescope (TMT) that consists of a
near-infrared (0.84 to 2.4 $\mu$m) imager and integral field spectrograph
(IFS). The IFS makes use of a lenslet array and slicer for spatial
sampling, which will be able to operate in 100's of different modes,
including a combination of four plate scales from 4 milliarcseconds (mas)
to 50 mas with a large range of filters and gratings. The imager will have
a field of view of 34$\times$34 arcsec$^{2}$ with a plate scale
of 4 mas with many selectable filters. We present the preliminary design of
the data reduction system (DRS) for IRIS that need to address all of these observing modes. Reduction of IRIS data will have unique challenges since it will provide real-time reduction and analysis of the imaging and spectroscopic data during observational sequences, as well as advanced post-processing algorithms. The DRS will support three basic modes of operation of IRIS; reducing data from the imager, the lenslet IFS, and slicer IFS. The DRS will be written in Python, making use of open-source astronomical packages available. In addition to real-time data reduction, the DRS will utilize real-time visualization tools, providing astronomers with up-to-date evaluation of the target acquisition and data quality. The quicklook suite will include visualization tools for 1D, 2D, and 3D raw and reduced images. We discuss the overall requirements of the DRS and visualization tools, as well as necessary calibration data to achieve optimal data quality in order to exploit science cases across all cosmic distance scales.

\end{abstract}

\keywords{integral field spectroscopy, data reduction pipeline}

\section{INTRODUCTION}
\label{sec:intro}  

IRIS~\cite{Larkin2010} (InfraRed Imaging Spectrograph) is the near-infrared
diffraction-limited imager and integral field spectrograph (IFS) for the
Thirty Meter Telescope (TMT).  The IRIS IFS will have four spatial scales
using a lenslet array (4 and 9 mas) and a slicer (25 and 50 mas). 
Ref~\citenum{Larkin2016}
shows the updated the FoV in each configuration of the lenslet and
slicer. In addition, IRIS will also have an 34$\times$34 arcsec$^{2}$
field of view (FoV) imager (including small chip gaps)
using four Hawaii-4RG (4096$\times$4096) infrared arrays. IRIS wavelength
coverage is 0.84-2.4 $\mu$m which currently plans to utilize a minimum of 42 different filters, encompassing broad and narrow-band filters. There are currently plans for at least 12 gratings that cover spectral resolving powers of R=4000, 8000 and 10000. With all of these configurations, IRIS will have hundreds of operating modes. This combined with observing with a multi-conjugate adaptive optics system on TMT will provide unique challenges to the data reduction system, which needs to provide real-time reduction allowing users to evaluate their data quality and orientation on target.

The planning and data reduction tools for instruments on next generation
ground-based telescopes (20 to 40-m class telescopes) will require a level of
sophistication similar to that of a spaced-based missions. The scale of these
telescopes requires large international partnerships in order to fund their
construction and operation, and will need to address a range of diverse users
and astronomical studies. Data from these instruments will be significantly
advanced, and  astronomical users will require planning software and data
reduction pipelines that work with a variety of science cases that deliver
publishable data products without requiring users to be instrument experts. In
addition, in order to optimize the time more efficiently, users will require
real-time feedback during their observations so they can adjust and make the
best use of their telescope time. In order to be a scientifically productive
instrument, it is necessary not only to have a fully developed pipeline that
has incorporated multiple science cases but one that give instantaneous
feedback that allow for exploratory science that benefits the classical mode
of observing and the dual imaging and spectroscopy modes offered by IRIS.

\section{Data Reduction System (DRS)}
\label{sec:drs}

The IRIS data reduction system (DRS) is planned to provide real-time
($\lesssim$ 2 minutes) data processing of imaging and spectroscopic data,
as well as a full off-line reduction package. The DRS will provide
visualization tools for raw and reduced data to facilitate data assessment
and analysis for real-time and off-line use. There will be three modes of
the reduction: imager data, lenslet IFS data, and slicer IFS data. However,
because both the lenslet IFS and slicer IFS share the same detector and
gratings, they will also share many of the same DRS algorithms. The IRIS
DRS is also responsible for processing all raw readouts from each science
detector (imager and spectrograph) and generating a raw science quality
frame.

The architecture of the full DRS package will be a pipeline, the model used
for many existing instruments.  In software, a pipeline is a chain of
processing elements (i.e. algorithms) arranged so that the output of each
element is the input of the next. The data reduction software \textit{system} will serve to link all the algorithms together and provides the necessary software infrastructure. All IRIS data reduction algorithms will be custom-designed for IRIS final data products. The basis of some 
algorithms will be adapted from previous IFS instruments, such as OSIRIS\cite{Krabbe2004}, GPI\cite{Marie2010,Perrin2014},
NIFS, and SINFONI\cite{Modigliani2007} pipelines.  Numerous
near-infrared imagers exist (e.g. NIRC2, NACO~\cite{Lenzen2003,Rousset2003}), and they will be
leveraged to provide algorithms for the Imaging mode whenever possible.

The IRIS DRS will be written in Python and data files will use the flexible
image transport system (FITS). Python is advantageous since it is open
source, free, has a large community of developers, easy to learn and use,
and portable. Many open-source packages are available that are applicable
for astronomical data reduction, such as NumPy, SciPy, Pandas, and Astropy.
Python has developed momentum in the Astronomical community with recent
pipelines such as HIPE\cite{Ott2010} (Herschel), CASA\cite{McMullin2007}
(ALMA and JVLA) and
MOSFIRE\footnote{https://keck-datareductionpipelines.github.io/MosfireDRP/}
(Keck). In addition, Python has been selected as the standard reduction
software packages developed at Space Science Telescope Institute for the
James Webb Space Telescope (JWST) as well as the Large Synoptic Survey
Telescope\cite{Juric2015} (LSST).

The IRIS DRS needs to be developed during the design and fabrication of the
instrument, since it needs to be fully delivered during the integration
phase of the IRIS imager, spectrograph, on-instrument-wavefront sensors
(OIWFS), and NFIRAOS at NRC-Herzberg. The DRS will be crucial for this
integration phase for commissioning each of these sub-components 2 years
before TMT+IRIS first light.

\subsection{Overview}

\begin{figure}[ht]
\begin{center}
\begin{tabular}{c}
\includegraphics[height=11cm]{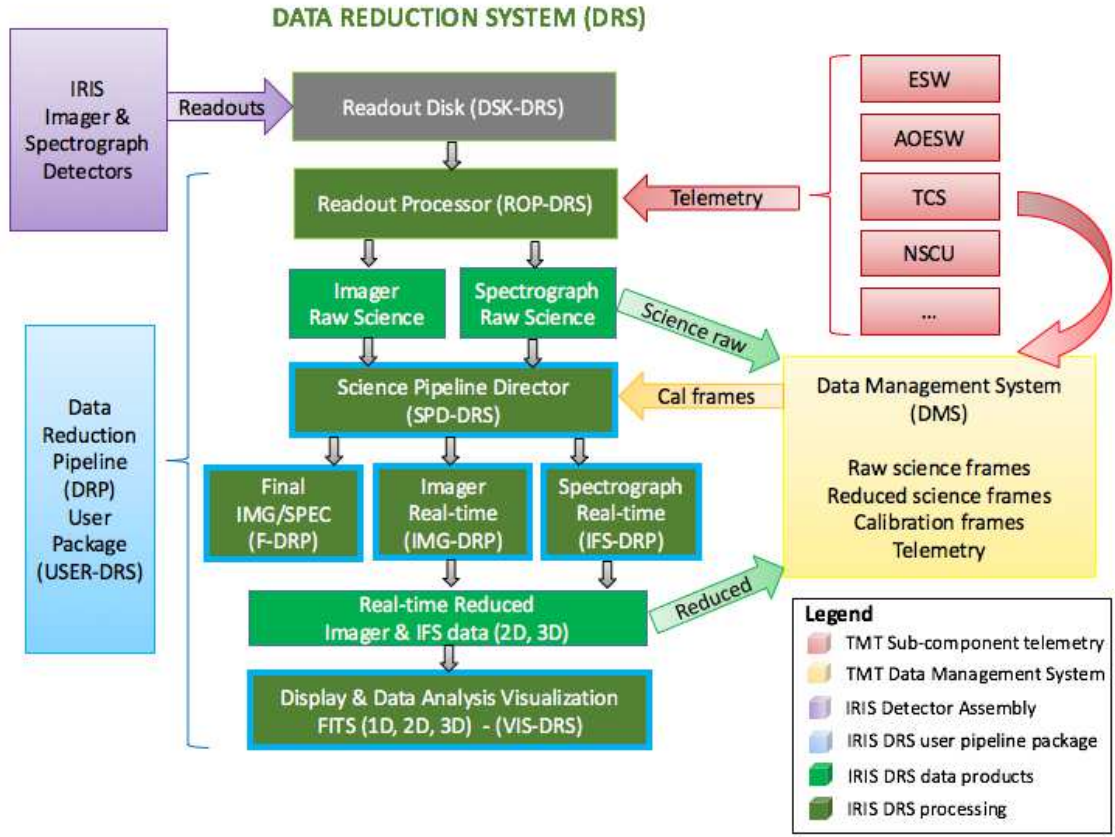}
\end{tabular}
\end{center}
\caption[example]
{ \label{fig:block}
The IRIS Data Reduction System (DRS) block diagram illustrates the flow of
data products, data processing, and data deliverable. The DRS is responsible
for storing individual raw readouts from the imager and spectrograph
detectors (purple) and is stored on a local disk (grey) for real-time data
processing (dark green). Real-time telemetry (red) from the executive
software (ESW), adaptive optics executive software (AOESW), TCS (Telescope
Control System), NFIRAOS Science Calibration Unit (NSCU), and other
sub-components will be retrieved by the DRS readout processor (ROP-DRS).
The ROP-DRS will process raw readouts from the detectors (e.g., up-the-ramp
sampling) to generate raw science frames. A real-time data reduction
pipeline for the imager (IMG-DRP) and spectrograph (IFS-DRP) will be
conducted on-sky, and final data reduction will be processed by the user
using the F-DRP and data pipeline package (USER-DRS; blue). Both raw
science frames and reduced science frames (2D, 3D) are then saved to the
TMT data management system (yellow) for archiving. A quicklook display and
data analysis visualization will be used during real-time and
post-processing.
}
\end{figure}

The IRIS DRS will need to coordinate all data processing from the
Hawaii-4RG detectors from the imager and spectrograph, and communicate with
the TMT data management system (DMS), TMT and AO execute software for all
necessary metadata for processing. Figure \ref{fig:block} shows the block
diagram of how the DRS will process all IRIS images and data, while
interacting with all other TMT and AO sub-components. 

We briefly describe the overall layout and flow of the IRIS DRS. Individual reads are readout from IRIS detectors from the imager and spectrograph to the readout disk (DSK-DRS).  These reads are immediately copied to the readout processor computer (ROP-DRS).  When an exposure is
complete, the ROP-DRS combines all of the reads into a single raw frame for the spectrograph and imager. During the raw frame creation process, the sampling scheme that has been selected by the user will be implemented (e.g. up-the-ramp sampling).  Finally, the FITS header is created by a process that polls telemetry data from the various instrument and telescope services such as: executive software (ESW); adaptive optics executive software (AOESW); telescope control system (TCS); and NFIRAOS Science Calibration Unit~\cite{Moon2010} (NSCU).  This telemetry provides critical
metadata that is needed by the DRS.  Once this step is complete, the raw
frames are copied to the data management system (DMS). The DMS will add the required observatory
header keywords to the FITS for proper archival following the Common
Archive Observation Model~\cite{Dowler2008,Dowler2012} (CAOM). The frames
containing all of the raw detector reads will be stored in an IRIS archive.  

The science pipeline director (SPD-DRS) pulls the necessary calibration files
(e.g., flats, white light scans, dark frames) from the DMS for data imager and
IFS processing.  The reduction algorithms are performed on the imager
real-time (IMG-DRP) and spectrograph real-time (IFS-DRP) pipelines. The
real-time pipeline will run a subset of routines from the final imager and
spectrograph reduction pipeline (F-DRP) in order to more efficiently provide
the user with reduced data. The F-DRP will will contain all algorithms that
any user can access to process off-line at their host institution(s). The
real-time reduced science frames (2D, 3D) are then saved to the DMS for
archiving. The raw and reduced science and calibration frames can be displayed
using a quicklook display and data analysis visualization (VIS-DRS) tools.
The final off-line data reduction pipeline (USER-DRS) includes the ROP-DRS,
SPD-DRS, F-DRP and VIS-DRS.  With the USER-DRS, the user can fully process and
visualize their data using the full set of routines.  

In the following sections, we will go over the DRS requirements (\S\ref{sec:req}), algorithms for each of the pipelines (\S\ref{sec:alg}), calibration files required (\S\ref{sec:calib}), metadata required (\S\ref{sec:metadata}), storage formats (\S\ref{sec:storage}) and visualization of the data (\S\ref{sec:vis}).


\subsection{Requirements}
\label{sec:req}

The TMT and IRIS high-level requirement on the IRIS DRS are the following:

\begin{itemize}
\item Reduced data for the imager (2D) and spectrograph (3D) shall be available for user assessment in less than 30 seconds and 60 seconds, respectively, to verify telescope pointing, instrument configuration, object geometry and data quality.  
\item The DRS will include a visualization software package for viewing and conducting basic analysis tools for 1D, 2D, and 3D FITS files.
\end{itemize}

\subsection{Algorithms}
\label{sec:alg}

The IRIS DRS will essentially have both real-time and final data pipelines
(DRP) that will address imager and IFS data. The DRP user package will include
the full set of algorithms. The imager DRP will include the following
algorithms: sky/dark subtraction; correction of detector artifacts (e.g.
crosstalk, bias adjustments); correction of cosmic rays; flat fielding; field
distortion correction; flux calibration; PSF calibration; and advanced shift
and add (mosaicking). The IFS DRP will include the following algorithms:
sky/dark subtraction; correction of detector artifacts; correction of cosmic
rays; flat fielding (slicer IFS only); spectral extraction (separate routines
for slicer IFU and lenslet IFU); wavelength solution (separate routines for
slicer IFU and lenslet IFU); cube (x, y, $\lambda$)  assembly (separate
routines for slicer IFU and lenslet IFU); and residual atmospheric dispersion
correction.
The real-time pipelines (IMG-DRP and IFS-DRP) will utilize the subset of the final pipeline (F-DRP).  This subset includes; sky/dark subtraction, correction of detector artifacts, correction of cosmic rays, flat fielding, spectral extraction, and cube assembly (x, y, $\lambda$).

Table \ref{tab:imaging} and \ref{tab:spectroscopy} summarize the algorithms used by the imager and spectrograph DRP. Table \ref{tab:advanced} summarize the advanced algorithms.  Each description includes a basic outline of what input data it receives, what other auxiliary data it requires (calibration data are described in Section \ref{sec:calib}), what it outputs, and what steps are performed to achieve its goal.

The IRIS spectrograph will have four plate scale modes; the lenslet (4mas and
9mas) and slicer (25 mas and 50 mas). The slicer mode will require an extra
flat-fielding step, which is standard for slit-type reductions. The
sky-subtraction will be performed by utilizing a scaled sky-subtraction that
accounts for changes in the absolute flux of OH lines, as well as variations
in specific vibration bands of the OH lines \cite{Davies2007}. Furthermore, we
are also investigating the use of more advanced sky-subtraction algorithms,
such as {\sc skycorr}~\cite{Noll2014}, which uses solar and atmospheric data
to model the sky emission model throughout a night of observations. This procedure was developed around Cerro Paranal site, however it can be utilized at other sites where a wealth of atmospheric data exist such as Mauna Kea.

\begin{table}[ht]
\caption{DRS imaging subroutines. Note: FRS = Frames, SCI = Science, CR =
Cosmic Ray, DIV = Division, DIST = Distortion}
\label{tab:imaging}
\begin{center}
\begin{tabular}{|l|l|l|l|}
\hline
DRS post processing             & Input                      & Output                    & Function \\
\hline
Generate master dark            & Darks                      & Master dark               & Median combine \\
Dark subtraction                & SCI                        & Dark subtracted FRS       & Subtraction \\
Remove detector artifacts       & Dark subtracted FRS        & Cleaned FRS               & Bad pixel and CR removal \\
Flat Fielding                   & Cleaned FRS                & Flat fielded FRS          & DIV by normalized flat field \\
Scaled sky-subtraction          & Flat fielded FRS           & Sky-subtracted SCI        & Scale factor from sky FRS \\
Field distortion correction     & Sky-subtracted SCI         & DIST corrected FRS  & Field distortion correction \\
Flux calibration                & DIST corrected FRS   & Flux calibrated SCI       & Flux calibration \\
Mosaic/Combine SCI              & Flux calibrated SCI        & Final combine SCI         & Dither shifts \\
\hline
\end{tabular}
\end{center}
\end{table}

\begin{table}[ht]
\caption{DRS spectroscopy subroutines. Note: * - Slicer only, ADC - Atmospheric Dispersion Correction}
\label{tab:spectroscopy}
\begin{center}
\begin{tabular}{|l|l|l|l|l|}
\hline
DRS post processing       & Input                   & Output                  & Function \\
\hline                                              
Generate master dark      & Darks                   & Master dark             & Median combine \\
Dark subtraction          & SCI                     & Dark subtracted FRS     & Subtraction \\
Remove detector artifacts & Dark subtracted FRS     & Cleaned FRS             & Bad pixel and CR removal \\
Flat Fielding*            & Cleaned FRS             & Flat fielded FRS        & DIV by normalized flat field \\
Spectral extraction       & Cleaned SCI             & Extracted SCI           & Advanced spectral extraction \\
Wavelength calibration    & Extracted SCI           & Wav calibrated SCI      & Least square minimization \\
Cube assembly             & Wav calibrated SCI      & 3D SCI cubes            & Cube assembly \\
Scaled sky-subtraction    & 3D SCI cubes            & Sky-subtracted SCI      & OH and continuum scaling \\
Residual ADC              & Sky-subtracted SCI      & ADC SCI                 & Atm. Dispersion Correction \\
Telluric correction       & ADC SCI                 & Telluric corrected SCI  & Telluric feature removal \\
Flux calibration          & Telluric corrected SCI  & Flux calibrated SCI     & Flux calibration \\
Mosaic/Combine SCI        & Flux calibrated SCI     & Final combined SCI      & Dither shifts \\
\hline
\end{tabular}
\end{center}
\end{table}

\begin{table}[ht]
\caption{DRS advanced post processing subroutines. Note: SCI = raw science frame, EXP = Exposure, PSF = Point spread function, PSF-R = PSF-reconstruction}
\label{tab:advanced}
\begin{center}
\begin{tabular}{|l|l|l|l|}
\hline
Advanced post processing  & Input                       & Output                      & Function \\
\hline
Optimizing readouts   & Readouts per EXP & Individual SCI  & S/N optimization \&\ telemetry \\
PSF-reconstruction        & Telemetry per SCI  & Generated PSF-R per SCI & PSF-R \\
\hline
\end{tabular}
\end{center}
\end{table}

\subsection{Calibration}
\label{sec:calib}

Auxiliary data used in DRP algorithms are called calibration data.  This includes both on-sky data (that is not of the astronomical target itself), daytime calibration frames, and other sub-component metadata.  Metadata is non-image information that will typically come from
the header of raw FITS files, or from IRIS, TCS, and/or NFIRAOS via
the observatory ESW event service. The NFIRAOS Science Calibration Unit
(NSCU) will include a calibration system that will facilitate the taking of
daytime calibration frames, such as arc lamp spectra, white light flat
field images, and pinhole grids for measuring distortion. Table \ref{tab:calib} summarizes the required calibration files necessary for the IRIS DRS.

\begin{table}[ht]
\caption{Calibration Frames. Note: * = SPEC only, PTG = pointing, D-Map = Distortion Map, Env = Environmental, DTC = Daytime calibration, NTC = Nightime calibration}
\label{tab:calib}
\begin{center}
\begin{tabular}{|l|l|l|l|l|}
\hline
Name                     & Reference Type & Source                & Algorithms \\
\hline
Atm. Dispersion Residual & Metadata       & IRIS ADC              & Atmospheric Correction \\
Arc lamp spectra*        & CAL (2D)       & IRIS DTC (NSCU)       & Wavelength solution \\
Bad pixel map            & CAL (2D)       & IRIS DTC              & Correction of detector artifacts \\
Dark Frame               & CAL (2D)       & IRIS DTC and NTC      & Dark subtraction \\
Env metadata             & Metadata       & ESW, FITS header      & All \\
Fiber image              & CAL (2D, 3D)   & IRIS DTC (NSCU)       & PSF Calibration \\
Flux calibration star    & CAL (2D, 3D)   & IRIS On-sky           & Extract Star, Remove Absorption Lines \\
Instrument config        & Metadata       & ESW, FITS header      & All \\
Lenslet scan*            & Rect Matrix,   & IRIS DTC (NSCU)       & Spectral Extraction \\
                         & CAL (2D)       &                       & \\
NFIRAOS config           & Metadata       & ESW, FITS header      & All \\
Pinhole Grid (D-Map)     & CAL (2D)       & IRIS DTC (NSCU)       & Field distortion correction \\
PSF metadata             & Metadata       & ESW, FITS header      & PSF calibration \\
PSF star                 & CAL (2D, 3D)   & IRIS on-sky           & PSF calibration \\
Sky frame                & CAL (2D, 3D)   & IRIS on-sky           & Sky-subtraction \\
Telescope config PTG     & Metadata       & ESW, FITS header      & All \\   
\hline
\end{tabular}
\end{center}
\end{table}

\subsection{Metadata}
\label{sec:metadata}

Metadata is additional information stored in the FITS file header that describes
essential data taken at the telescope, AO system, instrument, and all other sub-component systems. Metadata will relay information regarding individual observations or calibrations that are necessary to the DRS.  

\noindent The metadata written to the FITS headers by IRIS will serve two
purposes for the DRS:
\begin{itemize}
\item Provide necessary environmental information about the observation to
reduce the data.
\item Identity the file type (i.e. science, arc, flat, etc).
\end{itemize}

Metadata functions as a way to relay important environmental information
about the telescope, instrument and AO system to the DRS.  The metadata
written will be used for the majority of the DRS algorithms.  For example,
temperature and pressure information is needed to determine the position of
the spectra on the detector. The DRS will require metadata from the
following systems: telescope control system; tip-tilt and focus sensors
(e.g., wavefront sensor counts); deformable mirror status; pupil plane
location; AO telemetry; and PSF reconstruction. For example, the DRS will
need to know the temperature/pressure in the folowing locations: primary
mirror back-side, primary mirror front-side, secondary, outside the dome,
grating turret wheel, slicer array, dewar and detector (imager and
spectrograph), and NFIRAOS.  In addition, adopting the model in which every
read is saved, to make use of this mode it is important that real-time
telemetry information is stored within the FITS headers. This will be
critical information needed in order to correct for PSF variations from
read-to-read.

Metadata will be used by the DRS to identify which frames are science frames
(e.g, imager, IFS slicer, IFS lenslet), flats, darks and calibration frames.
Since the DRS will operate automatically when data is readout from the
detector the DRS needs to identify all frames necessary to reduce without any
ambiguity. The DRS will require a standardization of the metadata in order to
properly identify the frames.  The assignment of the metadata should be
automatically determined by the mode of the instrument (i.e. if the astronomer
needs to take darks, the frame will be identified as a dark). The assignment
of the metadata should be seemless to the astronomer observing with IRIS. This
work flow will mitigate the accidental mis-assignment of data type in each
FITS file.

\subsection{Storage Format Discussion}
\label{sec:storage}

Astronomy data formats have been using the FITS standard since the 1970s.
FITS consists of two portions; a binary portion which stores the data in the
form of an image/spectrum or table, and an ASCII portion which contains
keywords and values associated with the data.  The ASCII portion, or
FITS header stores metadata about each observational frame.

Metadata requirements of the DRS open up the considerations of
different storage formats with the use of IRIS.  This process is a
challenging task as the instrument team attempts to project what the standard
format will be at the time of TMT first light. For example, LSST is currently considering HDF5~\cite{hdf5} and ASDF~\cite{Greenfield2015} format instead of FITS, though their data rates far exceed any modern
telescope by many magnitudes. JWST on the other hand is considering ASDF
over FITS. While the FITS format is the standard
format used throughout observatories and space telescopes, it was created well before many well established on-line databases (e.g SQL\cite{Chamberlin1974}) and parsing code became an industry standard in the software community. 

One of the difficulties of the FITS format is the limitation and flexibility of the 8
character keyword used to identify values that may be needed by the DRS~\cite{Thomas2015}.
These header key words often become more like random strings and can become
less organized.  For example, when dealing with adaptive optics data, the DRS
will need to keep track of all wavefront sensor changes in the AO system, in order to accurately reconstruct
the PSF and process raw readouts. This keyword values will be rapidly changing and we may want a more flexible data storage system for data searching and parsing. One solution would be steer-away from the FITS standard, such as Hierarch used by ESO\footnote{http://fits.gsfc.nasa.gov/registry/hierarch\_keyword.html}, which adds multiple levels of key words and strings. This ``hierarchical" structure enables a cleaner organization as well keys that
are in a understandable language.  However, this solution is adhoc to the
FITS standard and is not yet globally accepted in the astronomical community. Thus far, TMT and IRIS has selected the FITS format to proceed, but our team is actively monitoring other data storage methods that may be implemented by LSST and JWST.

\subsection{Visualization}
\label{sec:vis}

The IRIS DRS will include visualization software for both real-time
observing and post-processing data analysis. Since the DRS will be running
real-time, IRIS users will require an easily functioning tool to display
and assess the quality of the raw and reduced imager and spectrograph data
sets. The DRS will require that the visualization tool will view 1D, 2D and
3D raw and reduced data.

There already exists well designed and easy-to-use visualization community
tools for displaying 2D images (e.g. DS9
SAOimage\footnote{http://ds9.si.edu}).  3D visualization tools on the other
hand are generally more specialized towards a specific instrument, and
historically have been  developed by the instrument team (e.g., OSIRIS,
GPI). A larger community effort is needed for these tools to reach the same
maturity as the 2D visualization tools. Development and advancement of a 3D
visualization tool is necessary for IRIS reduced data cubes. The 3D
visualization tools should include the ability to collapse the entire data
cube along different axises and functions (i.e., median, average). The
visualization tool should also allow easy inspection of individual spectra,
integrated spectra over a drag-able selection window. PSF assessment tools
will also need to be implemented for both the 2D and 3D data sets (e.g,
FWHM, Strehl ratio calculation). 

General visualization tools will be included: adjust brightness (stretch)
and contrast of the frames; image stretch schemes (e.g, linear, log, power
law); invert the stretch; pan; recenter; zoom in and zoom out; collapse
image to a specific channel; compute statistics on regions of pixels or
spaxels (e.g., sum, error, area, surface brightness, mean, median, minimum,
maximum, variance, standard deviation); centroid on the peak of a source;
plot horizontal, vertical and diagonal cuts across the image; display a
surface plot; and display a contour plot; and have functionality to
overplot multiple images and spectra with specified regions. Our team is
working on developing the full requirements of the visualization, that
advances current imaging functions, and identifies additional features that
are needed by the IRIS science team.

\section{Challenges}

The astrometric and photometric requirements for IRIS pose unique challenges for data processing of diffraction-limited imager and spectrograph for a 30m aperture. In this section, we highlight just a \textit{few} of the major challenges that the DRS will face in developing a pipeline that generates superb science-ready images and data cubes.

\subsection{Simultaneous observing modes}

IRIS is designed to have a 34$\times$34 arcsec$^{2}$ FoV imager with a pickoff mirror sending the central region (optimized NFIRAOS correction) to the IFS. This design allows for simultaneous data to be taken with the imager and IFS. This mode opens up a new avenue of science cases, but also poses some technical challenges for the DRS observing modes. In addition, the current design has a single filter wheel in the collimated beam before the pickoff mirror to the IFS, so filters selected need to work for both the imager and IFS. PSF-reconstruction will need to optimized for both the central field for the IFS as a function of wavelength, as well as optimized for the larger field of view of the imager. The simultaneous observations with the imager and IFS, also means that narrow- and broad-band spectrograph filters will need to be photometrically calibrated for the imager, as further described below.


\subsubsection{Filters}

The preliminary filter set for IRIS can be found in Ref.~\citenum{Larkin2010}. The IRIS filter set will be similar to OSIRIS filter data sets, and other near-infrared imaging cameras, which would include broadband and narrowband filters in Z, Y, J, H and K bands. While the exact throughputs of the filters will not be determined until the filters are procured, their wavelength bandpass (e.g.,5, 20\%) have consequences for the dual imaging and
spectroscopy modes.

The IRIS optical design has the NFIRAOS corrected beam feeding both the imager
and IFS. Along the optical path, the IRIS imager first samples the 34"x34"
FoV, and then a pick-off mirror feeds the optical-axis to the IFS. A large
filter wheel mechanism resides inside the imager optical train and feeds both
the imager and spectrograph.  This poses a challenge in the dual operation
mode as the spectroscopic filter set have different bandpasses than the
traditional near-infrared imaging filter sets.  This means that spectroscopy
driven observing programs will drive the filter used for parallel imaging. 
 
\begin{figure}[ht]
\begin{center}
\includegraphics[height=12cm]{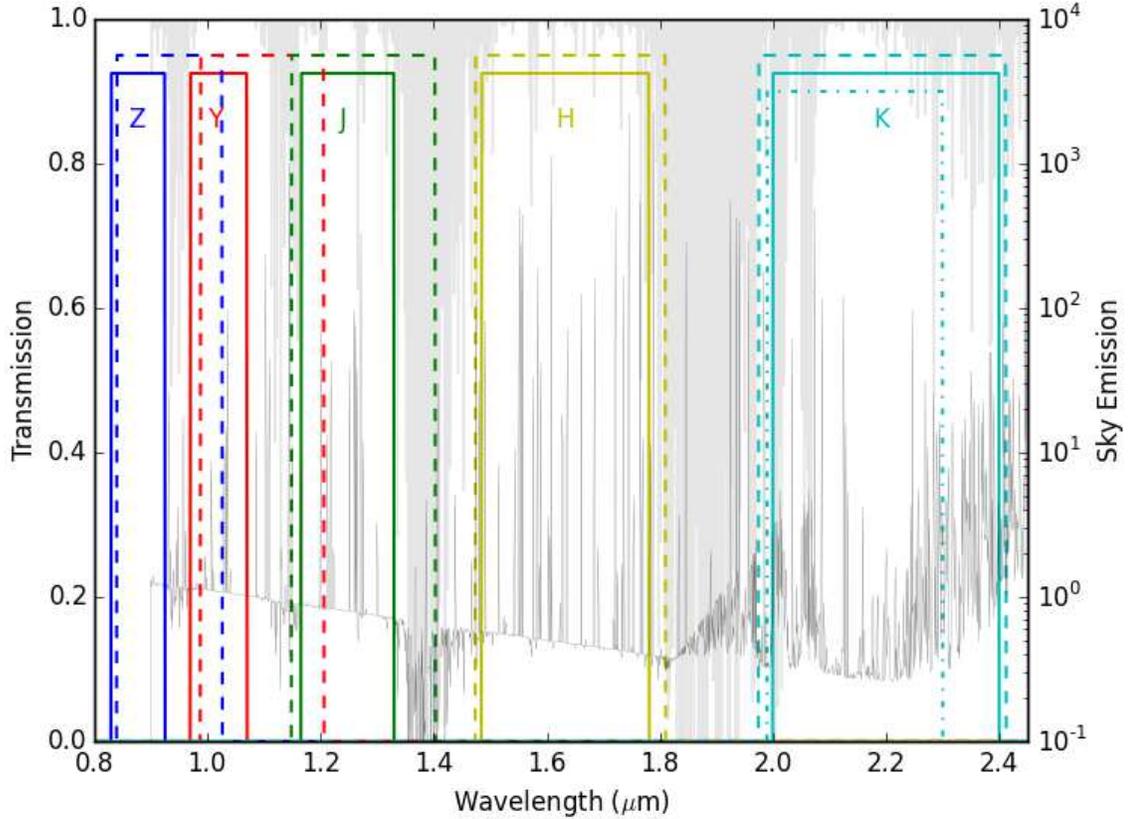}
\end{center}
\caption[example]
{ \label{fig:filters}
IRIS near-infrared broadband filters with the atmospheric transmission and sky
background (solid black line) taken on Mauna Kea (Gemini near-IR
transmission).
The light gray shows
opaque regions in the atmosphere.
The solid black lines shows the sky emission (Gemini near-IR sky
background). The solid lines are the
standard near-IR imaging filters (Z, Y, J, H and K). The dash lines are the
IRIS spectroscopic filters (Zbb, Ybb, Jbb, Hbb and Kbb) that 20\% bandpass.
The dash-dotted line is the standard Ks filter. The filters are staggered for
clarity since the total throughput per wavelength are not yet known.}
\end{figure}

There are two sets of spectroscopic filters; broadband and narrowband
filters.  The narrowband filters are 5\% bandpass to provide some wavelength coverage, while offering a larger FoV for the IFS modes. Spectroscopic broadband filters (i.e., Zbb, Ybb, Jbb, Hbb, Kbb) are designed to cover as much wavelength real estate as possible, and does not have to worry about atmospheric and backgrounds as traditional imaging filters (e.g., Z, J, Ks) were designed to optimize. Figure \ref{fig:filters} shows the broadband
filter sets for IRIS.  Table \ref{tab:filters} illustrates the transmission differences
between the spectroscopic broadband filters and the standard near-infrared filter bandpasses.
The broadband spectroscopic filters Hbb and Kbb are the most similar to the standard near-infrared equivalents.  While Zbb, Ybb and Jbb have more significant differences than
the traditional broadband imaging filters (Z, Y,, J).  Ybb not only shows the greatest
difference $>$2$\times$ but almost contains $\sim$3$\times$ more sky
emission. In addition, Zbb, Ybb and Jbb overlap with nearby filters whereas
the typical near-infrared equivalents do not. These different bandpasses will need to be fully characterize for the imaging camera, especially if a number of the science cases focuses primarily on using the spectroscopic filters.

\begin{table}[ht]
\caption{Transmission ratio between imager and spectrograph filter bandpasses}
\label{tab:filters}
\begin{center}
\begin{tabular}{|l|c|c|c|c|c|}
\hline
Ratio        & Zbb$/$Z & Ybb$/$Y & Jbb$/$J & Hbb$/$H & Kbb$/$K \\
\hline
Filter              & 1.96  & 2.18  & 1.55  & 1.14  & 1.09 \\
Filter + Atmosphere Transmission & ...   & 2.11  & 1.35  & 1.12  & 1.09 \\
Filter + Sky          & ...   & 2.83  & 1.29  & 1.11  & 1.09 \\
\hline
\end{tabular}
\end{center}
\end{table}

\subsubsection{Imaging Dithering}

The simultaneous imaging and spectroscopic modes may pose some challenges
in imaging and IFS data reductions. The IFS FoV is significantly smaller than the imager, and dither patterns between the imager and IFS will require careful planning and coordination. The construction of clean sky background that will be beneficial to both the imager and IFS will need to be particularly considered. The near-infrared sky is variable both due to thermal and OH-emission variability on minute time scales. For instance, long exposures ($>$ 600 sec) which are beneficial to dark-limited spectrograph observations would keep the imager data fixed at a stationary dither pattern. In contrast, when the imager science requires rapid dithering and telescope offsets this will impact the IFS total integration on source, as well as a collective sky frame(s) for background subtraction.

There are several sky cases encountered when constructing the sky background,
that depend on the density of objects observed. The following cases are examples of different dithering strategies:

\begin{enumerate}

\item object(s) fills the majority or all of the field of view (high
density; e.g. nebulae, dense star clusters) $\implies$ separate sky (large
dither) 

\item object(s) have medium density to low density (e.g. galaxy
clusters, open star clusters) $\implies$ science for sky (depending on the size of the objects; medium to small dithers) 

\item object(s) have medium to high density (e.g., stellar clusters, Galactic Center) but require small dither pattern for astrometric accuracy

\item object(s) have very low density $\implies$ science for sky (tiny dithers)

\end{enumerate}

For the low density object distributions, the tiny dither would be
sufficient to construct a sky background. One of the difficulties with this
method may be dealing with large CRs and hot pixels. The large density
object distributions would be similar to how other imagers deal with this
problem which is a large dither on blank sky to use for the sky background
construction.  Medium density object distributions could use two possible
solutions; (1) Sky dithers or (2) 2-target (N-target) dither.  Currently in
order to remove the sky background from the IFS, it is necessary to dither
off source for a few frames (1/3 time) to take a pure sky spectrum for use
in the scaled-sly algorithm (ref).  One possibility is to randomly dither
for the IFS sky which could be used to generate the imager sky.  The other
solution is to use a 2-target (or N-target) dither, where there are two IFS
targets within the FoV of the imager, it would be possible to dither to
both targets between exposures.  However, ideally more dithers would be
necessary for the masking process. Such a process could be accomplished
efficiently in a dense field where multiple targets require IFS
spectroscopy.

We are currently investigating these dither pattern scenarios, especially alternative methods for dealing with the sky background removal for both the imager and IFS.

\subsection{Data Rates and Storage}

One of the goals of TMT observatory's DMS is to archive raw data from all
TMT instruments for data storage and delivery to astronomical users. The
DMS is currently only being designed to store raw science quality files.
Our team is currently exploring the technical and scientific merits for
storing all individual reads from the imaging and spectrograph detectors,to
improve the quality of the reduction in both real-time and during
post-processing. Indeed, there are several missions that have greatly
benefited with saving all individual detector reads (e.g, NASA Spitzer)
that have generated additional science results that would not have been
possible without this capability.

Raw science-quality near-infrared astronomical frames are typically produced in the following:

\begin{itemize}
\item Reads are read from the detector into memory (readouts)
\item Intermediate reduction is performed for all reads for a user defined
exposure time
  \begin{itemize}
  \item Reference pixel subtraction, linearization, deinterlacing, bad pixel flagging 
  \item Up-the-ramp (UTR)/Multiple Correlated Double Sampling (MCDS)
computed for each coadd
  \item Coadds added/averages
  \end{itemize}

  \item A final file with combined readouts are written to disk as a FITS file (raw science frame)
\end{itemize}

In this typical readout method, the intermediate reduction is run before
the science-quality raw FITS file is created. Our team has been exploring
the possibility and advantages of saving every individual read to allow for
greater flexibility and performance improvements in the reduction process.
For example, if the \textit{seeing} degrades for a select number of
individual reads, those reads can be removed in post-processing to improve
the overall S/N of the raw science frame. This will require that metadata
from the AO system, TCS, and other supporting systems would need to be
recorded in real-time. The astronomical user then has the flexibility of
optimizing their reductions during post-processing using the needed
telemetry associated with each readout. This is especially true (and
likely) if there are newer sampling schemes or masking techniques not
considered yet by the instrument team, then the astronomical user can
greatly benefit from storing all individual reads.


Table \ref{tab:rates} explores the maximum data rates possible with the
imager and spectrograph setup. The maximum data rates calculated assumes
the following configuration: a pixel clocking rate of 400 MHz with 64
readout channels gives a minimum read time of 0.72 seconds; Correlated
Double Sampling (CDS) clock time of 2.16 seconds per exposure; 4k$\times$4k
detector will have 64 MB per exposure; and the observing run will be 16
hours per night (science with calibrations); and assuming 91 total
observing nights per year. This method is with (unrealistically) taking raw
frames continuously in CDS with the minimum integration time (including 1
reset) and ignoring overhead between frames (on average 1/2 read time and
could be as much as 1 read time). Table \ref{tab:rates}, column 2 shows the
maximum data rates is all individual readouts are saved. Each of these
calculations ignore real world overheads, such as resets, pausing exposures
during target acquisition, and telescope dithering. Since each individual
read are in integers, these files will be written with 16 bits/pixel.

The total data rates per detector and 5 detectors, including raw and
readout frames, is listed in Table \ref{tab:rates}, columns 3 and 4.
Currently, the planned implementation of data storage is that raw science
(FITS) frames are stored on the DMS. Individual readouts will be stored on
the IRIS stand-alone readout disk, and our team is investigating future
archiving and long-term solutions. To store $\sim$3 months of individual
readouts taken at the maximum data rates would require $\sim$280 TB storage.

\begin{table}[ht]
\caption{
 \label{tab:rates}
Maximum IRIS data rates with continuous data taking configuration. The total per detector is calculated by summing the raw frames and the readouts.}
\begin{center}
\begin{tabular}{|l|l|l|l|}
\hline
Raw Frames         & Readouts           & Total Per Detector  & Total Per 5 Detectors \\
\hline
29.6 MB/s          & 44 MB/s            & 73.6 MB/s           & 368 MB/s \\
104 GB/hr          & 156 GB/hr          & 260 GB/hr           & 1.27 TB/hr \\
1666 files/hr      & 5000 files/hr      & 6666 files/hr       & 33,330 files/hr \\
1.64 TB/night      & 2.44 TB/night      & 4.08 TB/night       & 20.4 TB/night \\
26,666 files/night & 80,000 files/night & 106,666 files/night & 533,330 files/night \\
149 TB/yr          & 222 TB/yr          & 371 TB/yr           & 1.81 PB/yr \\
2,426,600 files/yr & 7,280,000 files/yr & 9,706,600 files/yr  & 48,533,000 files/yr \\
\hline
\end{tabular}
\end{center}
\end{table}

\subsection{Spectral Extraction}

The IRIS DRS require separate and unique spectral extraction routines for the lenslet array and slicer IFS.

The spectral extraction methods used in the DRS for the slicer will be similar to Gemini NIFS and VLT SINFONI. The spectral format of the slicer will be spatially compact across the detector, and will be spaced uniformly across the detector. A traditional spectral tracing routine will work sufficiently for extracting individual spectra across the detector, utilizing optimal extraction\cite{Horne1986}.

The spectral extraction of the lenslet array IFS spectra will be an algorithm intensive  task for the DRS. Lenslet spectra are dispersed onto the detector with a staggered formation to avoid spectral overlap. In order to pack the lenslet spectra efficiently, spectra are usually minimally separated where the wings of the spectral PSF of each lenslet overlap with neighboring lenslets. The spacing between the lenslet spectra in the current design is $\sim$2.5 pixel on the detector. In order to reconstruct the spectra from the lenslet IFS, the extraction usually requires some form of deconvolution. 

In Keck OSIRIS, the implementation of the deconvolution is Richardson-Lucy algorithm, which assigns flux to each of the lenslets iteratively. The PSF of each lenslet is stable and is typically pre-determined by taking white light scans using a bright continuum source that measures the PSF of each lenslet accurately. This PSF template is then used in the iterative deconvolution routine to assign flux to each individual lenslet and the contribution to neighboring lenslets. While the deconvolution process for OSIRIS performs very well, there are several areas that could be improved and explored for newer routines that can take advantage of increased CPU power. For instance, we are exploring adopting new calibration methods to trace out the 2D PSF of each of the lenslet. Another possibility is using different deconvolution algorithms, such as Gauss-Seidel and Maximum-Entropy.

\section{Summary}

IRIS on TMT promises to be scientifically pivotal instrument providing the
diffraction limited near-IR spectroscopy and imaging.  The IRIS DRS will
play an important role in the acquisition and reduction of science data.
The goals of the DRS are to provide publishable science-quality data and
visualization tools for analysis and acquisition. In this paper we have
outlined the overall flow of the pipeline, as well as the types of
algorithms, calibrations, and metadata that it requires.  While FITS is the
format currently adopted by the instrument team, other formats are being
monitored for potential adoption in order to offer new capabilities and
provide a standard format with the rest of the Astronomical community. The
DRS will require visualization and basic analysis tools that can handle
multiple data dimensions (1d, 2d, and 3d). There are multiple challenges
associated with the DRS for satisfying the dual imaging and IFS modes and
data rates associated with saving individual readouts from the Hawaii-4RG
detector. The DRS will be an essential component of the IRIS instrument,
and will greatly aid in commissioning of TMT and NFIRAOS, as well as being
vital to the science productivity.

\acknowledgments

The TMT Project gratefully acknowledges the support of the TMT
collaborating institutions.  They are the California Institute of
Technology, the University of California, the National Astronomical
Observatory of Japan, the National Astronomical Observatories of China and
their consortium partners, the Department of Science and Technology of
India and their supported institutes, and the National Research Council of
Canada.  This work was supported as well by the Gordon and Betty Moore
Foundation, the Canada Foundation for Innovation, the Ontario Ministry of
Research and Innovation, the Natural Sciences and Engineering Research
Council of Canada, the British Columbia Knowledge Development Fund, the
Association of Canadian Universities for Research in Astronomy (ACURA) ,
the Association of Universities for Research in Astronomy (AURA), the U.S.
National Science Foundation, the National Institutes of Natural Sciences of
Japan, and the Department of Atomic Energy of India.

\bibliography{report} 
\bibliographystyle{spiebib} 

\end{document}